\documentclass[10pt]{article}
\usepackage{graphicx}
\graphicspath{{c/}}
\DeclareGraphicsExtensions{.eps,.ps}
\begin{document}
\vspace*{2.5cm}
{\Large\bf Quantum collapse of dust shells in 2+1 gravity}

\vspace*{0.8cm}
{\bf L. Ort{\'\i}z and M. P. Ryan, Jr.
\footnote{Instituto de Ciencias Nucleares, Universidad Nacional Aut\'onoma
de M\'exico, A. Postal 70-543, M\'exico 05249 D. F., Mexico}}

\vspace*{0.5cm}
\hspace*{1.5cm}\begin{minipage}[t]{9.5cm}\footnotesize{\it Abstract}

\hrulefill

This paper considers the quantum collapse of infinitesimally thin dust 
shells in 2 + 1 gravity.  In 2 + 1 gravity a shell is no longer a sphere, 
but a ring of matter. The classical equation of motion of such shells 
in terms of variables defined on the shell has been considered by 
Peleg and Steif \cite{stipel}, using the 2 + 1 version of the original 
formulation of Israel \cite{Isra}, and Cris\' ostomo and Olea \cite{crisol}, 
using canonical methods. The minisuperspace quantum problem can be reduced to
that of a harmonic oscillator in terms of the curvature radius of the shell,
which allows us to use well-known methods to find the motion of coherent wave 
packets that give the quantum collapse of the shell.  Classically,
as the radius of the shell falls below a certain point, a horizon forms. In the
quantum problem one can define various quantities that give ``indications'' of horizon
formation. Without a proper definition of a ``horizon'' in quantum gravity, 
these can be nothing but indications.  

\hrulefill
\end{minipage}
\vspace{1cm}
\section*{}

It is a great pleasure to be able to contribute to this special volume in
honor of Octavio Obregon.  One of us [M.R.] has known Octavio for more than 30 years
and has collaborated with him on many fruitful projects.  We both want to
wish him many more
productive years and continued success.

\section{Introduction}

This article describes a toy model of a toy model.  There has been some
interest over the years in minisuperspace quantization of thin shells as
models of the full quantum collapse of more complicated objects.  One of
most the vexing problems in this scenario has been whether an event horizon
will form and consequently some sort of ``quantum black hole,'' or whether
a shell of non-interacting particles will simply collapse to a point
where the uncertainty principle will provide a ``repulsive force'' and the
shell will reexpand into the same universe.

Simple considerations can answer this question up to a point.  It seems
obvious that for large enough masses one might expect a reasonably
peaked wave function centered on a radius near to but outside the
classical Schwarzschild radius of the shell would, as it moves toward
zero radius, maintain enough coherence to pass beyond the Schwarzschild
radius almost in its entirety and a ``black hole'' would form with high
probability.  For small masses one might expect that rapid spreading
would overwhelm the coherence of the wave function and the shell would
reexpand into our own universe with probability essentially (or exactly) one.

In full relativistic quantum gravity these simple ideas are fraught with
difficulties.  The most serious of these are:

1) The quantum problem becomes unphysical for masses much above the Planck
mass.  While the minisuperspace approach has frozen out all radiative modes
of the gravitational field and one is insisting on a single-particle
interpretation of the shell problem, the wave function in the Schr\" odinger
picture can still become pathological at a point where one might expect
graviton production to begin.  H\'aj{\'\i}\v cek \cite{haj1}
has given a sufficient
but not necessary condition that shows that we might expect problems
above a couple of Planck masses.  In \cite{cruz}, a qualitative
argument was presented that used Compton wavelength considerations to give a
similar bound.

2) There are very serious technical problems in the formulation of the
problem.  Many authors have used the full ADM method to construct a Hamiltonian for the system,
where they have chosen an internal time for the system in order to have a
true Hamiltonian
and a Schr\" odinger equation that allows the study of the time evolution of
the quantum system.  The choice of an internal time leads to the problem
of quantum formulations that are not unitarily equivalent.  Another problem
with these Hamiltonian formulations is that they often require an {\it ad
hoc\/} choice of a Hamiltonian in terms of variables defined on the shell
itself.  Several of these Hamiltonians have been given by various authors
\cite{haj1}, \cite{hkk}.  A Hamiltonian due to
H\'aj{\'\i}\v cek and Kucha\v r \cite{hyk}
has the advantage of being defined by a coherent procedure with no {\it ad
hoc\/} choices, and is formulated in terms of foliations of spacetime
by timelike surfaces.  I will discuss this Hamiltonian in more detail
below.  All of these Hamiltonians have limits on the mass of the shell
of a few
Planck masses.  One formulation that does not seem to
have a mass limit is based
on the Wheeler-DeWitt equation corresponding to one of the Hamiltonians
in \cite{hkk}.

3) Many of the Hamiltonians are quite complicated and there is no real
chance of finding analytic solutions to the Schr\" odinger equations of
these models.  Numerical solutions have been presented by a group consisting
of A. Corichi, G. Cruz, A. Minzoni. M. Rosenbaum  and M. Ryan of the UNAM, N.
Smyth of the University of Edinburgh, and T. Vukasinac of the University
of Michoacan \cite{cruz}.

In order to study a simpler problem (another toy model of a toy model),
Ryan \cite{vinart} considered the quantum collapse of a shell in Newtonian
gravity.  It is possible to derive a classical equation of motion for the
radius of the shell in terms of just the radius, $R(T)$, and the shell mass,
$M$.  This problem has the advantages that there is only one time and that
the equation is the same as that of a particle falling radially toward a
point particle of mass $M$ located at $r = 0$.  It is easy to quantize this
system, and the final Schr\" odinger equation is the analogue of the
$s$-state hydrogen-atom equation.  While exact analytic solutions exist,
one has to use scattering states, and the integrals needed to form wave
packets are not tabulated, so only approximate solutions were given in
\cite{vinart}.

The idea of the present article is to study the quantum collapse
problem in 2 + 1 gravity, where one can address some of the difficulties
mentioned above in a context where
some of the problems mentioned above do not exist. The quantum problem is
unambiguous for all masses, so there is no problem of wave function
pathologies.  As will be shown below, one possible Hamiltonian for this
problem
has the form of that of a harmonic oscillator. This will allow us to find
simple
analytic solutions that can be used to illustrate the development of the
quantum
collapse of the shell, and will allow us to investigate the problem of
horizon formation
in terms of analytic constructs. Our toy model of a toy model will help us
to understand
exactly what is happening in the numerical solutions that will be presented
in \cite{us}.
In the final section of the article we will mention possible extensions of the 2 + 1 
problem that might be the subject of future research.

The plan of the rest of the article is as follows.  Section 2 will be
a brief resum\'{e} of the literature on the general relativistic minisuperspace
problem.  Section 3 will set up the classical 2 + 1 problem,
Section 4 will consider the quantum problem and present solutions, while
Section 5 will be conclusions and suggestions for future study.

\section{The collapse of thin shells in relativistic quantum gravity}

The study of the quantum collapse of dust shells is about a decade old \cite
{strom}.  The classical collapse problem is fairly straightforward, and
can in principle be solved exactly. One assumes a $\delta$-function massive
(or null) shell where Birkhoff's theorem tells us that outside the shell
the metric is Schwarzschild and the metric inside the shell is Minkowski.
The Israel junction conditions can be used to derive the equation for the
evolution of the shell in terms of intrinsic variables on the shell itself,
the proper time, $\tau$, of an observer riding on the shell and the curvature
radius, $R(\tau)$, of the shell that he would measure.  The equation for the
motion
of the shell becomes \cite{Isra}
\begin{equation}
M = {\cal M}\left \{ 1 + \left (\frac{dR}{d\tau}\right )^2 \right \}^{1/2} -
\frac{{\cal M}^2}{2R}, \label{iseq}
\end{equation}
where ${\cal M}$ is the rest mass of the shell (a constant of motion) and
$M$ is the Schwarzschild mass of the exterior metric.  It is straightforward
to define $x = {\cal M} R$, $V \equiv dx/d\tau$ and $M/{\cal M}$ as our
Hamiltonian.  Assuming $V = V(P)$ and solving
$\partial H/\partial P = V(P)$ for the ``momentum" $P$, we find that $V =
\sinh^{-1} (P)$ and our Hamiltonian becomes
\begin{equation}
H = \cosh P - \frac{m}{2x},  \label{hajham}
\end{equation}
where $m = {\cal M}/M_{pl}$, $M_{pl}$ the Planck mass, a Hamiltonian
given by H\'aj{\'\i}\v cek \cite {haj1}. Unfortunately, this is not the only
Hamiltonian that gives the equation of motion (\ref{iseq}), and we are left
with the problem of defining an ``appropriate'' Hamiltonian for the
problem.  A number of Hamiltonians for different choices time, including
the time of an observer at the center of the shell where space is flat, and
a Wheeler-DeWitt equation identical to that for a relativistic charged
particle radially falling in a Coulomb potential are given
by H\'aj{\'\i}\v cek, Kay and Kucha\v r \cite{hkk}.

Kucha\v r and H\'aj{\'\i}\v cek \cite{hyk}, dissatisfied with such
{\it ad hoc\/} Hamiltonians,
have managed to construct a Hamiltonian for collapsing dust shells that
comes directly from an ADM reduction of the Hilbert-plus-matter action.
The problem with this approach is that, as spacetime quantities, the matter
variables are proportional to $\delta [R - R_0(\tau)]$, where $R_0 (\tau)$ is
the position of the shell as a function of shell proper time.  It is
virtually impossible to reduce the time derivatives of these delta functions
to reasonable variables in the matter Lagrangian that can give a shell
Hamiltonian that describes the motion purely in terms of canonical variables
on the shell.  Kucha\v r and H\'aj{\'\i}\v cek used an
ingenious method based on the fact that
the ADM reduction by a $3 + 1$ foliation is not restricted to foliation
by spacelike surfaces, but works just as well for foliations by timelike
surfaces.  Using this approach and a formulation of the dust fluid
velocity in terms of velocity potentials, they define a new Hamiltonian.
The cost of this consistent formulation is a very complicated Hamiltonian,
\begin{equation}
H= -\sqrt{2}R\left (1 - \frac{M}{R} - \sqrt{1 - \frac{2M}{R}}\cosh\frac{P}
{R}\right )^{1/2} \qquad R \geq 2M,            \label{karham1}
\end{equation}
\begin{equation}
H = -\sqrt{2}R\left (1 - \frac{M}{R} - \sqrt{\frac{2M}{R} - 1} \sinh
\frac{P}{R} \right )^{1/2} \qquad 0 \leq R \leq 2M.           \label{karham2}
\end{equation}

Almost all of the Hamiltonians that have been given
(except for the Wheeler-DeWitt equation of
 \cite{hkk}) seem to have mass limits beyond which the wave functions
become pathological.  These Hamiltonians are all so complicated that it
seems impossible to find analytic solutions to assist in the interpretation.  In \cite{cruz}, Corichi et
al. have given a series of numerical solutions that give the evolution of
wave functions that are initially sharply peaked over a radius near the classical
horizon, $R = 2M$, for the $\cosh P$ Hamiltonian (\ref{hajham}) and the
Hamiltonian given by Eqs. (\ref{karham1}-\ref{karham2}).  All of these solutions
show evolution of the peak toward $R = 0$ with a bounce caused by the
boundary conditions at $R = 0$ with the appearance of interference fringes as well as
a rapid spread of the wave packet.  The
Kucha\v r-H\'aj{\'\i}\v cek Hamiltonian has wave functions similar to those of
(\ref{hajham}), but with many rapid oscillations superimposed.

Since these Hamiltonians are self-adjoint, unitary evolution implies that
a peak formed from scattering states will always rebound to $R = \infty$.
One can ask whether this behavior means that all quantum collapse of this
sort implies a rebound into our own universe.  Since $\tau$ is proper time
on the shell and $R$ is also a shell variable, such questions can only
be answered by knowing the global quantum spacetime surrounding the shell.
In any case, the scenario of the shell observer is that he sees (begging
questions of quantum measurement and the reduction of
the wave packet) the shell
collapse to some point near $R = 0$, where uncertainty principle effects
change the classical equations of motion and the shell rebounds (actually,
a shell where the particles do not interact directly with one another
``passes through itself'' and reexpands, that is, each radially infalling
particle passes through $R = 0$ and the azimuthal angle $\theta$ jumps from
the initial $\theta_0$ to $\theta_0 + \pi$).  Even if the shell has
collapsed below its classical horizon, in finite proper time it will again
be above the horizon  and traveling toward $R = \infty$.  This
quasi-classical scenario is not surprising.  In proper time, a classical shell
that manages to avoid forming a curvature singularity at $R = 0$ would behave
in this way, but as $R$ becomes greater than $2M$ the shell would be in
a universe past our temporal infinity ($i_{+}$), or in ``another universe.''

This quasi-classical scenario is what one might expect to see for a large
mass where the quantum fluctuations would be small compared to the mass
and the evolution of the wave packet would be coherent long enough for
the shell to collapse past its horizon and the shell would emerge from
the horizon into a new universe. However, if the wave packet spreads
sufficiently so that the width is greater than the classical horizon radius,
we can see that a horizon might never form, and the shell would reexpand
into our own universe.

The problem of horizon formation in the quantum system is very difficult.
Event horizons are global features and one has to try to define a global
feature in a fluctuating manifold.  Of course, this quantum manifold must
be constructed in terms of the full minisuperspace canonical
quantum gravity of the shell-metric system.
In the shell case we tend to use some kind of approximation to construct
the spacetime metric.  Kucha\v r \cite{ku2} argues that for the simple
shell minisuperspace we may just replace the the shell mass (Schwarzschild
mass) and the shell radius in the metric outside the shell by the
corresponding operators to make a ``metric operator.'' The problem with
this metric operator is that it is a function of shell proper time, and
studies of the metric close to the shell \cite{tatj} cannot tell us
whether a true event horizon (tied to observer time at infinity) forms.

Other approximations are under study \cite{us}.  The simplest calculation
is to calculate $<\hat R(\tau)>$ and the uncertainty $\Delta R = \sqrt{<(\hat R - <\hat R>)^2>}$
and check whether $\Delta R$ becomes very large as $<\hat R>$ becomes small so that
$<\hat R>$ does not fall below the classical horizon and $\Delta R$ is larger than
 the classical horizon, which
can be taken as an indication of the non-formation of a horizon.  In 
\cite{us}, numerical evaluations of these two quantities will be presented
for the Hamiltonians (\ref{hajham}) and (\ref{karham1}-\ref{karham2}), and
they
suggest that no horizon forms for small masses.  Another possibility (to
be considered in \cite{us}) would be to take M$|\psi (R, \tau)|^2$ to be a
classical density $\rho (R, \tau)$ and calculate the classical metric due
to a classical fluid with this mass distribution and see whether a horizon
forms. Note that $M$ should be either the rest mass or the Schwarzschild
mass.  It is not yet clear which.  There are
technical problems with this calculation.  We have to
calculate the metric from a density that is given in terms of a solution
of the Schr\" odinger equation for our Hamiltonian, and there is no
guarantee that this density can be made to obey the equation
$T^{\mu \nu}_{\, \, \, \, \, ;\nu} = 0$ for our fluid. Another possibility
considered in
\cite{us} is that of the ``metric operator'' mentioned above, which was
extended to the whole
manifold outside the shell and used to define a ``quantum'' stress-energy
tensor.

The rotationless 2 + 1 problem has some advantages over the 3 + 1 problem.
The Hamiltonian can be constructed fairly easily, and, as will be shown,
has the form of a harmonic oscillator. The Schr\" odinger
equation for this Hamiltonian has well-known analytic solutions.  The
expectation value of $R$ and its uncertainty can, in principle, be
calculated analytically.  The analogue of the other calculation using $\rho
(R, \tau)$ is much simpler than in the 3 + 1 case. The classical horizon is
easily
found.  In the next section these ideas will be considered.

\section{The classical problem of shells in 2 + 1 gravity}

The first element we need for this problem is an equation for the radius of
the shell.  This problem has been studied in detail by Peleg and Steif 
\cite{stipel}, using the 2 + 1 version of the original  formulation of 
Israel \cite{Isra}, and Cris\' ostomo and Olea \cite{crisol}, using canonical methods. 
We will review the calculation of Peleg and Steif
(using the parametrization of the
metric of Cris\' ostomo and Olea).

The calculation we will use follows that of Israel who, as mentioned above,
studied the collapse of a shell in 3 + 1 gravity, represented by a 
delta-function sphere of dust of
radius $R(\tau)$.  As we have mentioned, in 2 + 1 gravity the shell is a
circle, i.e.
a ring of matter, also of radius $R(\tau)$. The metric of spacetime will be 
written in circular coordinates, where flat space is represented by the metric
\begin{equation}
ds^2 = -dt^2 + dr^2 + r^2 d\theta^2.
\end{equation}

The equation for a circle $R(\tau)$ is
\begin{equation}
^{(3)}r = R(\tau), \qquad ^{(3)}\theta = \theta, \qquad ^{(3)}t = t(\tau).
\end{equation}
We will use the notation $i, j = 1, 2, 3$ and $A, B = 1, 2$.  We will now need 
a set of coordinates $\xi_A$ on the circle, which we will take to be $\xi_A =
(\tau, \theta)$, and we will define two three-vectors on the circle by
$e^i_{(A)}
\equiv \partial x^i/\partial \xi_A$, where $x^i$ are the coordinates
$t, r, \theta$ given
above.  We have
\begin{eqnarray}
e^i_{(\tau)} = (\frac{dt}{d\tau}, \dot R, 0),\\
e^i_{(\theta)} = (0, 0, 1).
\end{eqnarray}
The metric on the circle is ($ds^2 = g_{AB} d\xi^A d\xi^B$)
\begin{equation}
ds^2 = -d\tau^2 + R^2(\tau) d\theta^2.
\end{equation}

We now have to calculate the three-dimensional metric due to the
ring of matter.
We have to solve the 2 + 1 Einstein equations (necessarily with cosmological 
constant $\Lambda$ to avoid completely flat solutions), 
\begin{equation}
R_{ij} - \frac{1}{2}g_{ij} R = -\Lambda g_{ij},
\end{equation} 
inside and outside the circle. Since there is a Birkhoff theorem in
2 + 1 gravity,
these metrics in vacuum will be static or stationary.  We will study
the static case,
where the matter has no angular momentum.  In this case the metric
has the form
\begin{equation}
ds^2 = -f(r)dt^2 + \frac{1}{f(r)}dr^2 + r^2 d\theta^2,
\end{equation}
and the well-known solution is
\begin{equation}
f = B - \Lambda r^2,
\end{equation}
$B$ a constant.  Since $\Lambda$ has units of inverse length, we will
write, as is common, $\Lambda = \pm 1/\ell^2$.
The final form of the static, circularly-symmetric metric is
\begin{equation}
ds^2 = - \left( 1 - 2M \mp \frac{r^2}{\ell^2}\right ) dt^2 + \frac{1}{\left ( 1 - 2M \mp \frac{r^2}{\ell^2}\right )}
dr^2 + r^2 d\theta^2, \label{metrc}
\end{equation}
where we have taken $B$, following Cri\' ostomo and Olea, to be $1 - 2M$,
M a Schwarzschild mass.  Notice that this is not the choice of
Ba\~ nados, Teitelboim and Zanelli (BTZ) \cite{btz},
who take $B = -M_{\rm BTZ}$,
but we choose $1 - 2M$ following Crist\'o somo and Olea because we will want
the metric inside the ring to be the 2 + 1 AdS metric. These two choices of $B$
correspond to a shift in the zero of the mass parameter
\cite{BTZ2}, \cite{carl}.  The relation between these two masses (and others)
and the ADM mass is discussed in the Appendix.
This choice requires that the mass term be negative in order to have a 2 + 1
black hole with a horizon.  Notice also that we must have
$M > 1/2$ in order to have a horizon.  As in the 3 + 1 case, we
expect that outside the ring we will have a black hole metric with some
``Schwarzschild mass" $M$, that is,
\begin{equation}
ds^2 = - \left( 1 - 2M + \frac{r^2}{\ell^2}\right ) dt^2 + \frac{1}{\left ( 1 - 2M + \frac{r^2}{\ell^2}\right )}
dr^2 + r^2 d\theta^2, \label{metfin}
\end{equation}
and inside the ring we will have a metric that is as near as possible to
flat space (but with a cosmological constant).  This will be (\ref{metfin})
with $M = 0$, that is the 2 + 1 AdS metric,
\begin{equation}
ds^2 = - \left( 1 + \frac{r^2}{\ell^2}\right ) dt^2 + \frac{1}{\left ( 1 + \frac{r^2}{\ell^2}\right )}
dr^2 + r^2 d\theta^2.
\end{equation}

With these preliminaries we can follow the steps of Israel's derivation
of the equation of
motion of a shell in 3 + 1 gravity, a manner similar to that
used by Peleg and Steif \cite{stipel}.  For any
three-dimensional vector $A_A$, we can define components of on the
surface of the circle as
\begin{equation}
A_A \equiv A_i e^i_{(A)}, \qquad A^i \equiv A^A e^i_{(A)}.
\end{equation}
We will also need the vector $n^i$, the unit normal to the circle $R(\tau)$,
with $n^i n_i = +1$.
We can now use the standard equation
\begin{equation}
\frac{\partial A^i}{\partial \xi^A} = A^B_{\,\,\,\, ;A} e^i_{(B)} - A^B
K_{BA} n^i,
\end{equation}
where $K_{AB}$ is the extrinsic curvature of the ring, to arrive at the
equation
\begin{equation}
\frac{\partial u^i}{\partial s}{\big |}^{\pm} = \frac{\partial u^i}{\partial
\xi^A}
\frac{d\xi^A}{ds} = -U^A K^{\pm}_{AB} u^B n^i,
\end{equation}
or
\begin{equation}
n_i u^i_{\,\,\, ; j}u^j = -u^A u^B K_{AB} {\big |}^{\pm},\label{nueq}
\end{equation}
where $\pm$ means the quantity calculated for $r > R(\tau)$ (+) and
$r < R(\tau)$ (-). In order to calculate the left-hand-side of (\ref{nueq})
we will need expressions for $u^i$ and $n_i$.  Following
Israel, $u^r = \dot R$, $u^{\theta} = 0$ and $u_i u^i = -1$ gives us $u^t
\equiv X = \sqrt{f + \dot R^2}/f$.  There are many ways to find $n_i$, and
we obtain $n_t = -\dot R$, $n_r = X$ and $n_{\theta} = 0$.  We will also
need the definition,
\begin{equation}
\gamma_{AB} \equiv K^+_{AB} - K^-_{AB},
\end{equation}
and the Lanczos relation,
\begin{equation}
\gamma_{AB} - g_{AB}\gamma = 8\pi S_{AB}, \label{lanc}
\end{equation}
where $g_{AB}$ are the components of the induced metric on the surface in
terms of $\tau$ and $\theta$,
$\gamma = g^{AB}\gamma_{AB}$ and $S_{AB}$ is the surface stress-energy
tensor.  In our case, we will
be interested in a dust shell, so we will take $S_{AB} = \sigma u_A u_B$,
$\sigma$ the rest mass density
on the ring.

Following Cris\' ostomo and Olea, the Einstein equation for $G_{ij}n^i n^j$
becomes
\begin{equation}
^2R + K_{AB}^{\pm}K^{AB}_{\pm} - K^2_{\pm} = -2G_{ij}n^i n^j =
-\Lambda g_{ij}n^i n^j = -\Lambda.
\end{equation}
Subtracting the minus equation from the plus equation, and defining
$\tilde K_{AB} \equiv \frac{1}{2}
(K^+_{AB} + K_{AB}^-)$, we arrive at
\begin{equation}
\gamma_{AB} \tilde K^{AB} - \tilde K \gamma = 0,
\end{equation}
where $K = g_{AB} K^{AB}$.  The Lanczos relation gives us $S_{AB}\tilde
K^{AB} = 0$ or, since $S_{AB} = \sigma u^A u^B$,
\begin{equation}
u^A u^B \tilde K_{AB} = 0.
\end{equation}

Israel uses Eq.(\ref{nueq}), calculating the left-hand-side directly, and,
doing the same for our case,
we find
\begin{equation}
(n_i u^i_{\,\, ;j}u^j)^{\pm} = \frac{1}{f^{\pm} X^{\pm}} \left( \ddot R +
\frac{R}{\ell^2}\right ).
\end{equation}
Now, summing the plus and minus versions, and using $u^A u^B \tilde K_{AB} =
0$, we find
\begin{equation}
\left (\ddot R + \frac{R}{\ell^2}\right )\left [\frac{1}{\sqrt{f^+ +
\dot R^2}} + \frac{1}{\sqrt{f^-+ \dot R^2}}\right ] = 0.\label{nueqplus}
\end{equation}
Using $\gamma_{AB} = 8\pi(S_{AB} - g_{AB} S)$ and calculating $\gamma_{AB}
u^A u^B = u^A u^B(S_{AB} -
g_{AB}S = \sigma - \sigma = 0$, we find that  
\begin{equation}
\left (\ddot R + \frac{R}{\ell^2}\right )\left [\frac{1}{\sqrt{f^+ +
\dot R^2}} - \frac{1}{\sqrt{f^-+ \dot R^2}}\right ] = 0, \label{nueqmin}
\end{equation}
which is consistent.  Unfortunately, Israel uses the 3 + 1 equivalent of
(\ref{nueqmin}), where the right-hand-side is {\it not} zero, to find a
second-order equation of motion for $R(\tau)$ in terms of the rest mass
density, but our equations no longer contain any reference to this variable.  However, we can use, following Peleg and Steif \cite{stipel}, the Lanczos relation, Eq.(\ref{lanc}), directly. If we impose
the condition $\ddot R +  R/\ell^2 = 0$, we have that
$(n_i u^i_{\,\, ;j}u^j)^{\pm}$ are both zero,
so from Eq. (\ref{nueq}), $u^A u^B K^{\pm}_{AB} = 0$.  From $u_A
\equiv u_i e^i_{(A)}$, we find that
$u^A = (1, 0)$, so $K^{\pm}_{\tau \tau} = 0$. Now, from the Lanczos relation
we have that $\gamma_{\tau \tau} - g_{\tau \tau}\gamma = 8\pi S_{\tau \tau}$, but $\gamma_{\tau \tau} = 0$ and $g_{\tau \tau} = -1$,
so $S_{\tau \tau} = \sigma$, and finally, 
\begin{equation}
\gamma = 8\pi \sigma.
\end{equation}
Since $\gamma = g^{AB} \gamma_{AB} = -\gamma_{\tau \tau} +
(1/R^2)\gamma_{\theta \theta} = (1/R^2)\gamma_{\theta \theta}$, we have
\begin{equation}
\gamma_{\theta \theta} = 8\pi R^2 \sigma. \label{gamtt}
\end{equation}
We can now calculate $K_{\theta \theta}$ directly from its definition,
$K_{\theta \theta} = n_{\theta, \theta} - ^{(3)}\Gamma^k_{\theta \theta}
{\big |}_{r = R}$. Calculating the necessary $\Gamma^i_{jk}$ and using
the fact that $n_{\theta, \theta} = 0$, we find that $K_{\theta \theta} =
-fXR$, so
\begin{equation}
\gamma_{\theta \theta} = (fX)^- R - (fX)^+ R.
\end{equation}
From $S^{AB}_{\,\,\, ;B} = 0$, we can show that $\sigma = m/2\pi R$,
$m$ the total rest mass of the ring.

If we now consider the equation of motion of $R$, $\ddot R = -R/\ell^2$,
we see that it has a first
integral,
\begin{equation}
E = \frac{1}{2} \dot R^2 + \frac{R^2}{2\ell^2}.\label{ener} \label{eeq}
\end{equation}
Using $(fX)^- R - (fX)^+ R = 4mR,$ and inserting (\ref{ener}) in this
relation, we find
\begin{equation}
\sqrt{1 + 2E} - \sqrt{1 - 2M + 2E} = 4m,
\end{equation}
which can be solved for E as
\begin{equation}
E = \frac{M^2}{2(4m)^2} + \frac{M}{2} + 2m^2 - \frac{1}{2} = \frac{1}{32}
\left (\frac{M}{m} + 8m\right )^2 - \frac{1}{2}.
\label{eeq2}
\end{equation}

We would like to use Equation (\ref{eeq}) to construct a Hamiltonian
formulation of the problem.
This equation is simply the energy equation for a harmonic oscillator,
so it is obvious that
if we take $E$ as our Hamiltonian we can define a
momentum $P_R$ as $\dot R$, and we have
\begin{equation}
H = \frac{1}{2}P_R^2 + \frac{R^2}{2\ell^2}.
\end{equation}
Hamilton's equations for this Hamiltonian are equivalent to the equation
of motion $\ddot R  = -R/\ell^2$.

\section{Quantum collapse in 2 + 1 gravity}

\subsection{Schr\" odinger equation and evolution of wave packets}

Once we have a Hamiltonian in terms of the ring variables $R$ and $\tau$,
we can try to construct
a Schr\" odinger equation for the problem. Before we do this, we have to
study the units of our
variables.  If we write the classical relation
\begin{equation}
\left (\frac{M}{4m} + 2m \right )^2 - 1 = \dot R^2 + \frac{R^2}{\ell^2},
\label{hameq1}
\end{equation}
we see that the quantities on the right-hand-side have no units, so the
left-hand-side must also be
dimensionless, so $M$ and $m$ must be dimensionless.  We would like to
multiply the true
mass by some combination of $G$ and $c$ that will give a dimensionless mass.  We have to be careful about the law of gravitation in two space dimensions.  In general, Gauss's law would give
us, in $d$ dimensions, $F = GMm/r^{d-1}$, and in two space dimensions
$F = GMm/r$. This means that $G$ would have units of $[M]^{-1} [L]^2
[T]^{-2}$, and $G/c^2$ has units of $1/[M]$, and $GM/c^2$ is dimensionless.
If we consider the metric function $f(r) = 1 - 2M + R^2/\ell^2$, which
has to be
dimensionless, this means that M in conventional units is $GM/c^2$.
It is usual (see Ref. \cite{carl}) to define a ``Planck mass'' $M_{\rm Pl} =
c^2/G$ and a ``Planck length'' $L_{\rm Pl} = \hbar G/c^3$, even though there
is no particular reason to identify this mass and length with the 3 + 1
Planck mass and Planck length, especially if we have no knowledge of what
$G$ should be in two space dimensions.

In our case, $M$ means $M/M_{\rm Pl}$ and $m$ means $m/M_{\rm Pl}$.
In conventional units, our equation for $R$ becomes
\begin{equation}
\left (\frac{M}{4m} + \frac{2m}{M_{\rm Pl}}\right )^2 - 1 = \frac{1}{c^2}
\left (\frac{dR}{d\tau} \right )^2 +
\frac{R^2}{\ell^2}.
\end{equation}
Our ``Hamiltonian'' should have the units of energy for a conventional
quantum Hamiltonian, so
we should multiply Eq. (\ref{hameq1}) above by $M_{\rm Pl} c^2$, our
$R$-momentum $P_R$ becomes
$M_{\rm Pl} \dot R$ , and with
\begin{equation}
H = \frac{M_{\rm Pl}c^2}{2} \left (\frac{M}{4m} \right )^2 +
\frac{Mc^2}{2} + 2mc^2 \left (\frac{m}{M_{\rm Pl}}\right )^2 -
\frac{M_{\rm Pl}c^2}{2}, \label{hamval}
\end{equation}
and
\begin{equation}
H = \frac{P^2_R}{2M_{\rm Pl}} + \frac{M_{\rm Pl}}{2}\omega_0^2 R^2,
\end{equation}
with $\omega_0 = c/\ell$.

We want to quantize this system with $H$ replaced by an operator $\hat H$.
If we look at (\ref{hamval}), we see that the right-hand-side must become an
operator, so the left-hand-side must also be an operator. One way to achieve
this is to take $M$ to be a $q$-number, and $m$ and $M_{\rm Pl}$ $c$-numbers.
Of course, there is no special reason to make this choice, but we do so to
make contact with previous work (see, for example, \cite{ku2}).  If we
consider Eq. (\ref{hajham}), this choice is motivated by the fact that
the Schr\"odinger equation for (\ref{hajham}) is similar to the hydrogen atom
Schr\"odinger equation, with $m$ playing the role of $e^2$, and it is usual in the
hydrogen atom to take $e^2$ as a $c$-number rather than a $q$-number.

We can now take the wave function of the system to be $\tilde \Psi =
\Psi_M \Psi (R, \tau)$, with $\Psi_M (M)$ an approximate eigenstate of
$\hat M$ with eigenvalue $M_0$.  An exact eigenstate of $\hat M$ of this
type would be $\delta (M - M_0)$, but to avoid problems with the integral of
the square of a delta function we will assume that $\Psi_M$ is an
extremely sharply peaked wave function centered on $M = M_0$.  In this case,
our Schr\" odinger equation becomes ($\hat M^2 \Psi_M \approx M_0^2 \Psi_M$,
and realizing $\hat P_R$ as $-i\hbar \partial/\partial R$)
\begin{eqnarray}
\left [  \frac{M_{\rm Pl}c^2}{2} \left (\frac{M_0}{4m} \right )^2 +
\frac{M_0 c^2}{2} + 2mc^2 \left (\frac{m}{M_{\rm Pl}}\right )^2 -
\frac{M_{\rm Pl}c^2}{2}\right ] = \nonumber \\
= i\hbar \frac{\partial \Psi(R, \tau)}{\partial \tau} = -\frac{\hbar^2}
{2M_{\rm Pl}} \frac{\partial^2\Psi (R, \tau)}{\partial R^2} +
\frac{M_{\rm Pl}}{2}\omega_0^2 R^2 \Psi (R, \tau).
\label{schreq}
\end{eqnarray}
One curious fact about this equation is that, if we define $\Psi =
\exp(-iE\tau/\hbar)\psi(R)$, then the energy eigenvalues are
\begin{equation}
E_n = (n + \frac{1}{2})\hbar \omega_0 = (n + \frac{1}{2})
\frac{\hbar c}{\ell},
\end{equation}
and since $E$ is given by the first line of (\ref{schreq}), there is a
discrete relation between the Schwarzschild mass, $M_0$, and the rest mass,
$m$,
\begin{equation}
\left (\frac{M_0}{4m} + \frac{2m}{M_{\rm Pl}}\right )^2 - 1 =
(2n + 1)\frac{\hbar G}{c^3 \ell} = (2n + 1)\frac{L_{\rm Pl}}{\ell}.
\end{equation}
We will now return to units where $G = c = \hbar = 1$, $M_0$ now meaning
$M_0/M_{\rm Pl}$, $m$ now meaning $m/M_{\rm Pl}$, and $\ell$ meaning
$\ell/L_{\rm Pl}$. If we solve for $M_0$ in terms of $m$ and $\ell$, we find
\begin{equation}
M_0 = 8m^2\left [\sqrt{\frac{1}{4m^2} + \frac{1}{2m^2}\left[
\frac{1}{\ell}\left (n + \frac{1}{2}\right )\right ]} - 1\right ].\label{meq}
\end{equation}

We now want to write our Schr\" odinger equation in terms of dimensionless
variables.  For a moment we will return to conventional units, and define
the following dimensionless variables.  We define $y = R/\sqrt{\ell
L_{\rm Pl}}$ and
$T = c\tau/\ell$. The Schr\" odinger equation now becomes
\begin{equation}
i\frac{\partial \Psi (y, T)}{\partial T} = -\frac{\partial^2 \Psi (y, T)}
{\partial y^2} + y^2 \Psi(y, T)
\end{equation}
which has eigensolutions
\begin{equation}
\Psi = e^{-i(n + \frac{1}{2})T}\frac{1}{\sqrt{2^n n!}(\pi)^{1/4}}
e^{-y^2/2}H_n (y),
\end{equation}
$H_n$ Hermite polynomials.

Now, in order to make contact with previous work, we would like to study
the evolution of a wave packet sharply peaked around a value of $y = y_0$,
some distance outside the point where a classical horizon would form if the
radius of the shell were to fall below $R_H = \ell \sqrt{2M_0 - 1}$ and
follow its movement as the packet falls toward $R = 0$.
In previous work it was necessary to solve this problem numerically, since
in the 3 + 1 case analytic solutions to the Schr\" odinger equation did not
exist \cite{cruz} \cite{us}, and in the Newtonian case, \cite{vinart}, while
analytic solutions existed, it was impossible to sum the expansions for the
relevant wave functions to give an analytic expression for the collapsing
wave packet.  In the 2 + 1 case, however, we can give an exact analytic
expression for the wave packet as a coherent harmonic-oscillator state.
Even though the eigensolutions are nothing but harmonic oscillator wave
functions, the radial variable $R$ cannot be negative.  In order to keep
this from happening we will take the potential to be that of a half
oscillator with an infinitely hard wall at $R = 0$.  This potential is
shown in Figure 1.  This will mean that $\Psi (0, \tau)$ will always be
zero.  This boundary condition can be enforced by only expanding in odd-$n$
harmonic oscillator eigenfunctions. We will want to begin with a difference
of two Gaussian states, one peaked around $y = +y_0$ and the other around
$y = -y_0$, so that their sum at $y = 0$ is zero.
This state is
\begin{equation}
\Psi (y, 0) = \alpha [e^{-\frac{1}{2}(y - y_0)^2} - e^{-\frac{1}{2}(y +
y_0)^2}],\label{init}
\end{equation}
only valid for $y > 0$, with $\alpha$ a normalization constant.  Since this
is a sum of two Gaussian states, we can use standard techniques to construct
the difference between two coherent Gaussian states with (\ref{init}) as an
initial condition.  The result is
\begin{eqnarray}
\Psi(y,\tau) =  \alpha e^{-i\omega_0 \tau/2} e^{i y_0^2 \sin 2\omega_0
\tau/4} [\exp(-\frac{1}{2}\{ (y - y_0\cos \omega_0 \tau )^2\})
\exp (-iy y_0 \sin \omega_0 \tau )- \nonumber \\
 - \exp(-\frac{1}{2}\{ (y + y_0\cos \omega_0 \tau)^2 \} ) \exp (iy y_0
 \sin \omega_0 \tau) ], \label{cohstat}
\end{eqnarray}
which is zero at $y = 0$ for all $\tau$.  The normalization $\alpha$ is
easily found to be
$\alpha^2 = [\sqrt{\ell}\sqrt{\pi}(1 - e^{-y_0^2})]^{-1}$.

We would like to connect the variables that describe the wave function with
the radius of the
classical horizon, $R_H$, shown in Fig. 1.  We will make the following
definitions, $y_0 =
\lambda \sqrt{\ell} \sqrt{2M_0 - 1}$, $y = w\sqrt{\ell}\sqrt{2M_0 - 1}$,
$T = c\tau/\ell$.  If we take $\sqrt{\ell} = 1/\sqrt{2M_0 - 1}$, the
unnormalized probability density, $\rho \equiv \Psi^{*} \Psi/\alpha^2$,
\begin{eqnarray}
\rho = e^{-(w - \lambda \cos T)^2} + e^{(w + \lambda \cos T)^2}- \nonumber \\
-2e^{-w^2} e^{-\lambda^2 \cos^2 T} \cos(2\lambda w \sin T), \label{wpsi}
\end{eqnarray} 
is shown, for $\lambda = 3$, as a function of $w$ for various values of
$T$ in Figures 2-7.
We begin with a Gaussian packet at $T = 0$ that collapses toward $w = 0$,
developing interference fringes as it reaches a minimum for some
$w < \lambda$, then rebounds toward $w = \lambda$ again.
This pattern then repeats forever.  In order to compare our results with
earlier work, we will only be interested in one cycle of this pattern.
      
\subsection{The formation of a horizon}

In previous articles the quantization of the shell collapse was used to
study the possibility of the formation of
a horizon in the quantum collapse.  As mentioned in Sec. 2, this concept has
many difficulties.  An event horizon is a global construct and it has no
local definition.  This means that in quantum gravity one would have to
return to the starting point and try to define what a ``quantum horizon''
might be.  Once this definition has been decided upon, one must try to find
out if some collapse process will result in the formation of such a horizon,
with the result being a probability of
horizon formation.  Theories of quantum gravity in their present state are
far from being able to give us this result,
so in shell collapse some articles \cite{cruz}, \cite{us}, \cite{vinart}
have tried to give an estimate of horizon formation by finding out if a
sharply peaked wave packet, during its collapse toward $R = 0$, will fall,
in some sense, below the classical horizon radius, $R_H$.  In some sense,
because the packet will usually spread and basically will never lie entirely
below $R = R_H$. One could use the integral of $\Psi^{*} \Psi$ from $R = 0$
to $R = R_H$, which is a number less than one, as the probability of horizon
formation.  Another possibility would be to use the operator $\hat M$ in the
expression $\hat R_H = \ell\sqrt{2\hat M - 1}$ to define a
``horizon operator'' that could be used to define the probability of horizon
formation.  In previous work it was decided to use $<\hat R>$ as a function
of $\tau$, which will fall from $R_0$ to a minimum and then begin to
increase again. If this minimum is below the classical horizon, one
can say that a horizon forms and if not, not.  This is a yes-no answer
instead of a probability, but it is
a quick estimate.  We will use this concept below.

It is not difficult to calculate $<\hat R>(\tau)$ from the wave packet given
in (\ref{cohstat}), but the result is a complicated function that contains
the error function and Dawson's function, so trying to find the minimum of
$<\hat R>(\tau)$ by finding the point where $d<\hat R>/d\tau = 0$ requires
the solution of a transcendental algebraic equation, making it difficult to
give an analytic expression for the point where a horizon would form.
Instead we will use
the fact that, as can be seen from Fig. 5, that at $T = \pi/2$, the point
where the packet begins to rebound, the peak nearest $R = 0$ is high and
narrow.  We can use the position of this peak as a parameter to tell us
whether a horizon forms or not.  If the position of the peak is below $R_H$,
a horizon forms, and if not, not. In previous work it was found that for
large masses the shell collapse was so rapid that the wave packet fell below
$R_H$ so quickly that quantum mechanics did not allow it to rebound before
that point, while for small masses the rebound occurred for $R > R_H$.
In order to calculate the position of the maximum of $\rho$ nearest to
$w = 0$, we write $\rho$ as
\begin{equation}
\rho (y, T) = 2e^{-y^2_0 - y_0^2 \cos^2 T}[\cosh (2y y_0\cos T) - \cos(2yy_0
\sin T)].
\end{equation}
The packet is closest to $y = 0$ when $T = \pi/2$, which gives
\begin{equation}
\rho(y, \pi/2) = 2e^{-y^2}[1 - \cos (2yy_0)] = 4e^{-y^2} \sin^2 (yy_0).
\end{equation}
This function has zeros at $y = n\pi/y_0$, and either peaks or valleys
(or inflection points) at $d\rho (y, \pi/2)/dy = 0$, or
\begin{equation}
8e^{-y^2} \sin (yy_0)[-y\sin(yy_0) + y_0 \cos(yy_0)] = 0.
\end{equation}
An obvious solution to this equation is $y = n\pi/y_0$ for the zeros.  Another
possibilty is $y = y_c$, where
\begin{equation}
y_c = y_0 \cot (y_c y_0),
\end{equation}
which are peaks (it is easy to show that $d^2\rho/dy^2 < 0$).  If we define
$y_c = \gamma \sqrt{\ell} \sqrt{2M_0 - 1}$, we have
\begin{equation}
\gamma = \lambda \cot [\gamma \lambda \ell (2M_0 - 1)].
\end{equation}
for $\lambda = 3$ we can solve for $M_0$ as
\begin{equation}
M_0 = \frac{1}{2} \left [ \frac{1}{3\gamma \ell} \cot^{-1} \left
(\frac{\gamma}{3} \right ) + 1 \right ]. \label{coteq}
\end{equation}

We now have two relations, (\ref{coteq}) and $(M_0/4m + 2m)^2 - 1 = E$.  The
minimum for $M_0$ [the right-hand-side of (\ref{coteq}) decreases
monotonically with $\gamma$] is when $\gamma = 3$, since we want $y_c \leq
y_0$.  The numerical value of this minimum depends on $\ell$, and there
is no simple reason to choose any special value for $\ell$.  We can use
the relation between $M_0$, $m$ and $E$ to find another value for $M_0$ as
a function of $\ell$, and equating these two, we get a numerical value for
$\ell$.  There are various ways to define $M_0$ as a function of $\ell$ from
the energy relation.  If we consider the classical equation
\begin{equation}
\left (\frac{M_0}{4m} + 2m \right )^2 - 1 = \ell p_y^2 + \frac{y^2}{\ell},
\end{equation}
and take $y_0$ to be where $p_y = 0$, then $(M_0/4m + 2m)^2 - 1 = y_0^2\ell
= 18M_0 - 9$.  One solution of this quadratic equation is
\begin{equation}
M_0 = 8m^2 \left [17 + \sqrt{288 - \frac{2}{m^2}}\right ].
\end{equation}
This is real for $18m^2 > 1/18$, which gives a minimum for $M_0$ equal to
$M_0 = 17/18$, and for $\gamma = 3$,
\begin{equation}
M_0 = \frac{1}{2} \left (\frac{\pi}{36\ell} + 1 \right ) = \frac{17}{18},
\end{equation}
or $\ell = \pi/32 \approx 0.098$, and $M_0 (\gamma = 3) = 17/18 \approx
0.944$.

If we use the expectation value of $\hat H$ as our energy, we have $<\hat H>
= (M_0/4m + 2m)^2 - 1$, and calculate $<\hat H>$ to be
\begin{equation}
<\hat H> = \frac{1}{2} [1 + y_0^2(1 + e^{-y_0^2})(1 - e^{-y_0^2})^{-1}],
\end{equation}
in principle we can solve (using $y_0 = 9\ell[2M_0 - 1]$) for $M_0$, but
the transcendental equation is difficult to solve for $M_0$ as a function
of $m$ and $\ell$.  If we were able to assume that $y_0$ were sufficiently
large that the the factor $(1 + e^{-y_0^2})(1 - e^{-y_0^2})^{-1}$ were one,
we could solve the resulting quadratic for $M_0$ easily. However, for the
minimum $M_0$ from (\ref{coteq}) with $\gamma = 3$, we find that $y_0^2 =
\pi/4$, and our factor is almost equal to three.  We can find an approximate
solution by inserting $y_0^2 = \pi/4$ in the factor and write the quadratic
equation
\begin{equation}
\left (\frac{M_0}{4m} + 2m \right ) - 1 = \frac{1}{\ell} + (18M_0 - 9)
(1 + e^{-\pi/4})(1 - e^{-\pi/4})^{-1},
\end{equation}
which gives, using $\beta \equiv (1 + e^{-\pi/4})(1 - e^{-\pi/4})^{-1} = \coth
(\pi/8) \approx 2.68$
\begin{equation}
M_0 = 8m^2 \left (18\beta - 1 + \sqrt{(18 \beta)^2 - 36 \beta - \frac{1}{4m^2}
\left [(9\beta - 1) - \frac{1}{\ell}\right ]}\right ), \label{mbeta}
\end{equation}
which has real solutions for
\begin{equation}
4m^2 > \frac{(9\beta - 1) -1/\ell}{(18\beta)^2 - 36\beta},
\end{equation}
which has a minimum at $4m^2 = [(9\beta - 1) - 1/\ell]/[(18\beta)^2 -36\beta]$,
and inserting this expression in (\ref{mbeta}), and equating the result to
(\ref{coteq}) with $\gamma = 3$, we find $\ell (1/9\beta -1)[\pi\beta/4 +
(18\beta - 1)/(9\beta - 1)] \approx 0.180$.  This gives us $M_0(\gamma = 3)
= 0.742$.                                     

For the classical approximation we now have $M_0(\gamma = 1) = 2.62$,
while for $<\hat H>$ we have $M_0(\gamma = 1) = 1.66$. Since $<\hat H>$ should
give an answer closer to the correct one, we can take it as giving the best
value of $\ell$.  This means that for
\begin{equation}
0.742 < M_0 < 1.66,
\end{equation}
no horizon forms, while for $M_0 > 1.66$ one does.  For example, for $\gamma
= 1/2$, $M_0 = 2.80$.

From the Appendix, we can give these masses in terms of $M_{\rm BTZ}$, and
for
\begin{equation}
1.48 < M_{\rm BTZ} < 3.32,
\end{equation}
no horizon forms, while for $M_{\rm BTZ} > 3.32$ one does. 
                                             
Of course, as has been mentioned above, the rebound of the wave packet,
once it has passed the horizon does not mean that the shell returns through
the horizon into the same spacetime where it began. In spite of the fact
that in terms of proper time the shell exits the horizon in a finite time,
in terms of the time, $t$, of an observer at infinity this exit occurs at
time {\it after} $t = \infty$, or into ``another universe.''

\section{Conclusions}

The 2 + 1 calculation given above has the advantage of being the only model so far that has given analytic solutions to the quantum collapse problem.  This is a great advantage in understanding the origin of features in the numerical solutions given before in Refs. \cite{cruz}, \cite{us}, \cite{vinart}, especially the the existence of the interference fringes that were noticed in Refs. \cite{cruz} and \cite{us}.  The results for horizon formation are at least consistent with those of previous work.

Another possible model problem would be to study a simple relativistic model of gravity
such as the Nordstr\" om theory.  In a such a scalar theory the reduction
to a Hamiltonian formulation is relatively straightforward, but one of us (L. O.) (see Ref. \cite{lo}) has shown that the Hamiltonian is more complicated than that of ordinary 3 + 1 gravity. 
He has also shown that linearized gravity gives the same Hamiltonian as full gravity.

Still another possibility would be to carry out the program suggested in Ref. \cite{us} in the 2 + 1 case, that is, to try to define an average metric by writing the metric outside the ring in terms of coordinates $\xi$ (where $r = R(\tau) + \xi$) and $t = \tau$.  In this metric we can replace $R(\tau)$ by $\hat R$, to form a metric operator $\hat g_{ij}$ and calculate $<\hat g_{ij}>(\xi, \tau)$, where, for example, $\hat g_{\theta \theta} = (\hat R + \xi)^2$, so
\begin{equation}
<\hat g_{\theta \theta}> = <\hat R^2>(\tau) + 2<\hat R(\tau)> \xi + \xi^2,
\end{equation}
and use the classical Einstein tensor, $G_{ij}$, calculated from $<\hat g_{ij}>$ as $8\pi T_{ij}$ to define a classical stress-energy tensor $T_{ij}(\xi, \tau)$ (which automatically satisfies $T^{ij}_{\,\,\,\,\, ;j} = 0$).  From this $T_{ij}$, we can define, for example, a shell density
as a function of $\xi$ and $\tau$, which could be plotted. 

\section*{Appendix}

The mass parameter $M$ that appears in the black hole metric (\ref{metfin}) 
can be associated with the ADM mass of the black hole.  The ADM mass for the metric
with $B = -M$ is defined as \cite{BTZ2}, \cite{carl}
\begin{equation}
M_{\rm ADM} = \frac{M_{\rm Pl}}{16\pi} \int dr\, d\theta \, \sqrt{^{(2)}g} ^{(2)}R 
\approx \frac{\ell}{r} \frac{M_{\rm Pl}}{8} M_{\rm BTZ},
\end{equation}
where $^{(2)}g_{AB}$ is the two-metric on surfaces of constant $t$, and $^{(2)}R$ is the
Ricci scalar on these surfaces.  The factor $\ell/r$ is related to proper time at $r 
\rightarrow \infty$ \cite{carl} and is usually ignored, so we find that
\begin{equation}
M_{\rm BTZ} = \frac{8}{M_{\rm Pl}} M_{\rm ADM}.
\end{equation}
We can choose our mass unit to be any multiple of $M_{\rm Pl}$. In \cite{BTZ2} 
and \cite{carl} they take it to be $8M_{\rm Pl}$ while in \cite{stipel} it is taken to be
simply $M_{\rm Pl}$.

If we calculate $M_{\rm ADM}$ for our metric, we find that
\begin{equation}
M_{\rm ADM} = \frac{4}{M_{\rm Pl}} \frac{\ell}{r} (M - \frac{1}{2}) + A,
\end{equation}
where $A$ is a constant taken to be $(M_{\rm Pl}/4)r/\ell$ in \cite{BTZ2} and \cite{carl} that represents the 
zero of the ADM energy, $M_{\rm ADM} c^2$.  In older work \cite{btz}, BTZ chose $A = 
(M_{\rm Pl}/4)r/\ell + (M_{\rm Pl}/8)\ell/r$
as we have in this article.  The difference lies in the metric for $M = 0$.  In our case,
$M = 0$ is AdS, while for later BTZ work $M = 0$ represents
\begin{equation}
ds^2 = -\frac{r^2}{\ell^2} dt^2 + \frac{\ell^2}{r^2} dr^2 + r^2 d\theta^2.
\end{equation}

In our case, the choice of reference mass equal to $M_{\rm Pl}$ leads to a relation
between our $M$ and $M_{\rm BTZ}$, 
\begin{equation}
M = \frac{1}{2} M_{\rm BTZ}.
\end{equation} 
  
\section*{Acknowledgments}

We would like to thank A. Minzoni, and M. Rosenbaum
for many stimulating discussions.

\vfill\eject

\begin{figure}
\includegraphics[width=5in]{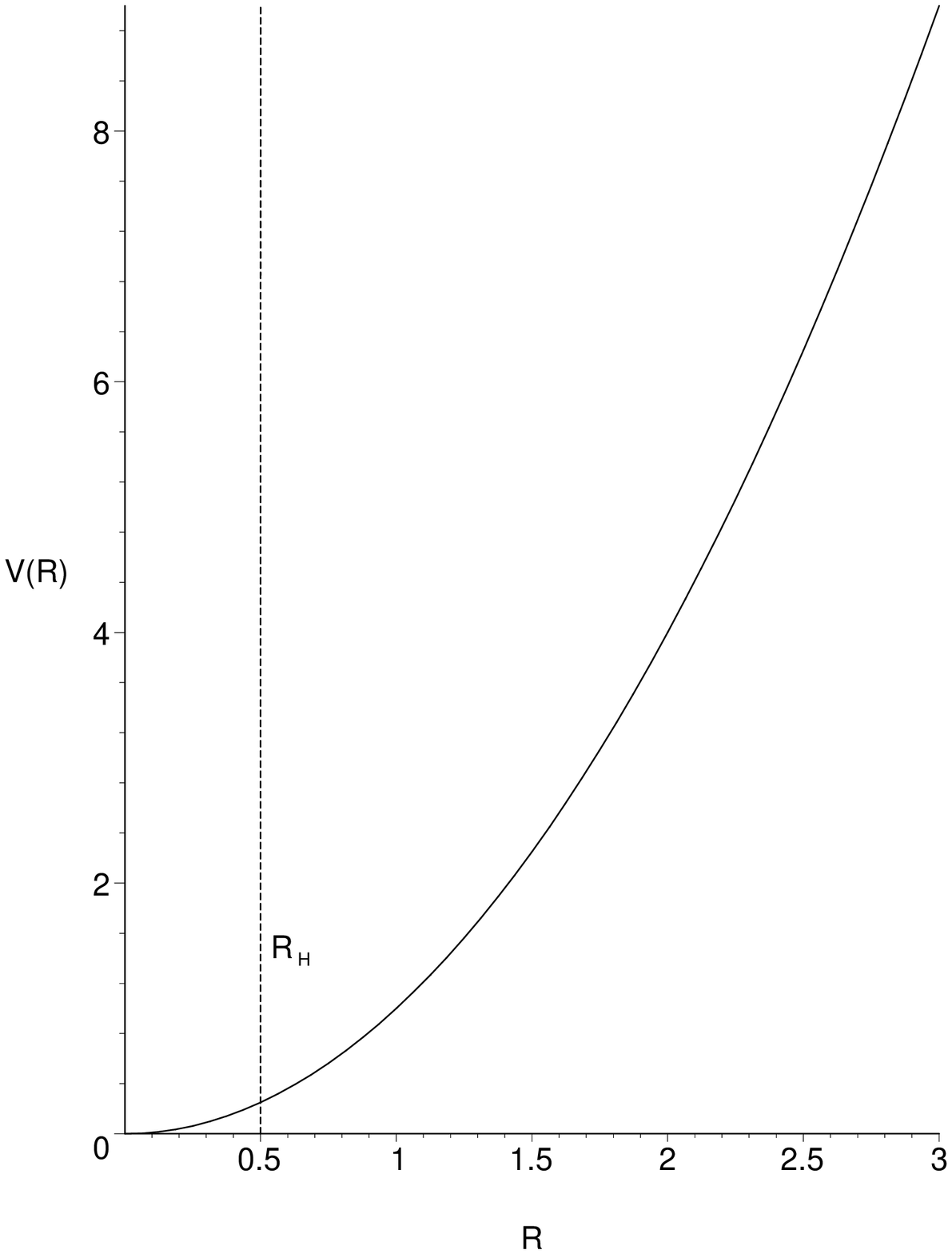}
\caption{
The harmonic oscillator potential for our problem with $M_{\rm Pl} \omega_0^2/2 = 1$.  The dashed line shows a typical position of the classical horizon radius, $R_H$.}
\end{figure}
\vskip 10 pt

\begin{figure}
\includegraphics[width=5in]{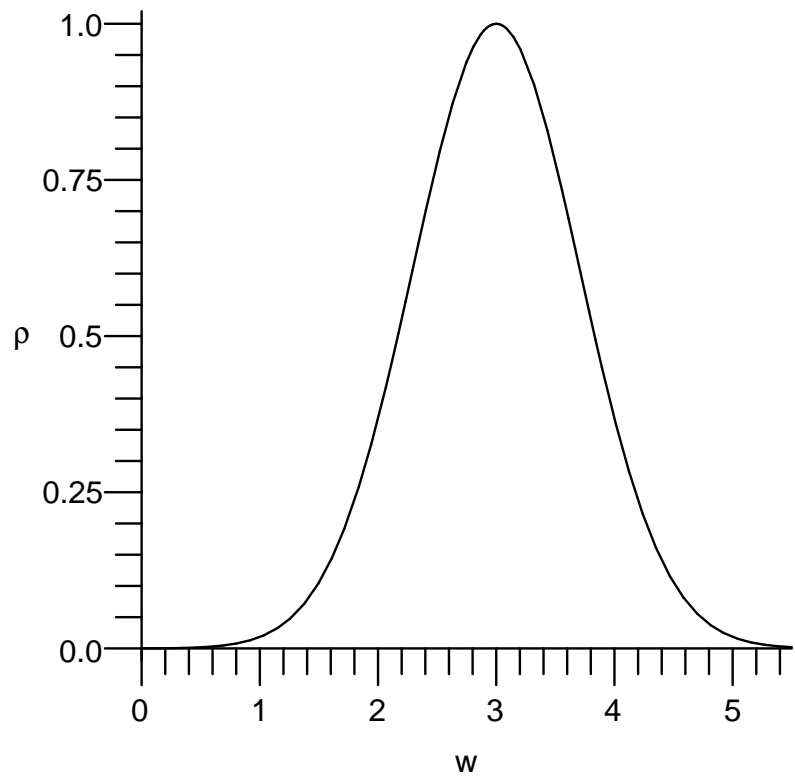}
\caption{
The probability density for the wave packet given in Eq. (\ref{wpsi}) for $T = 0$ (the initial state).}
\end{figure}
\vskip 10 pt

\begin{figure}
\includegraphics[width=5in]{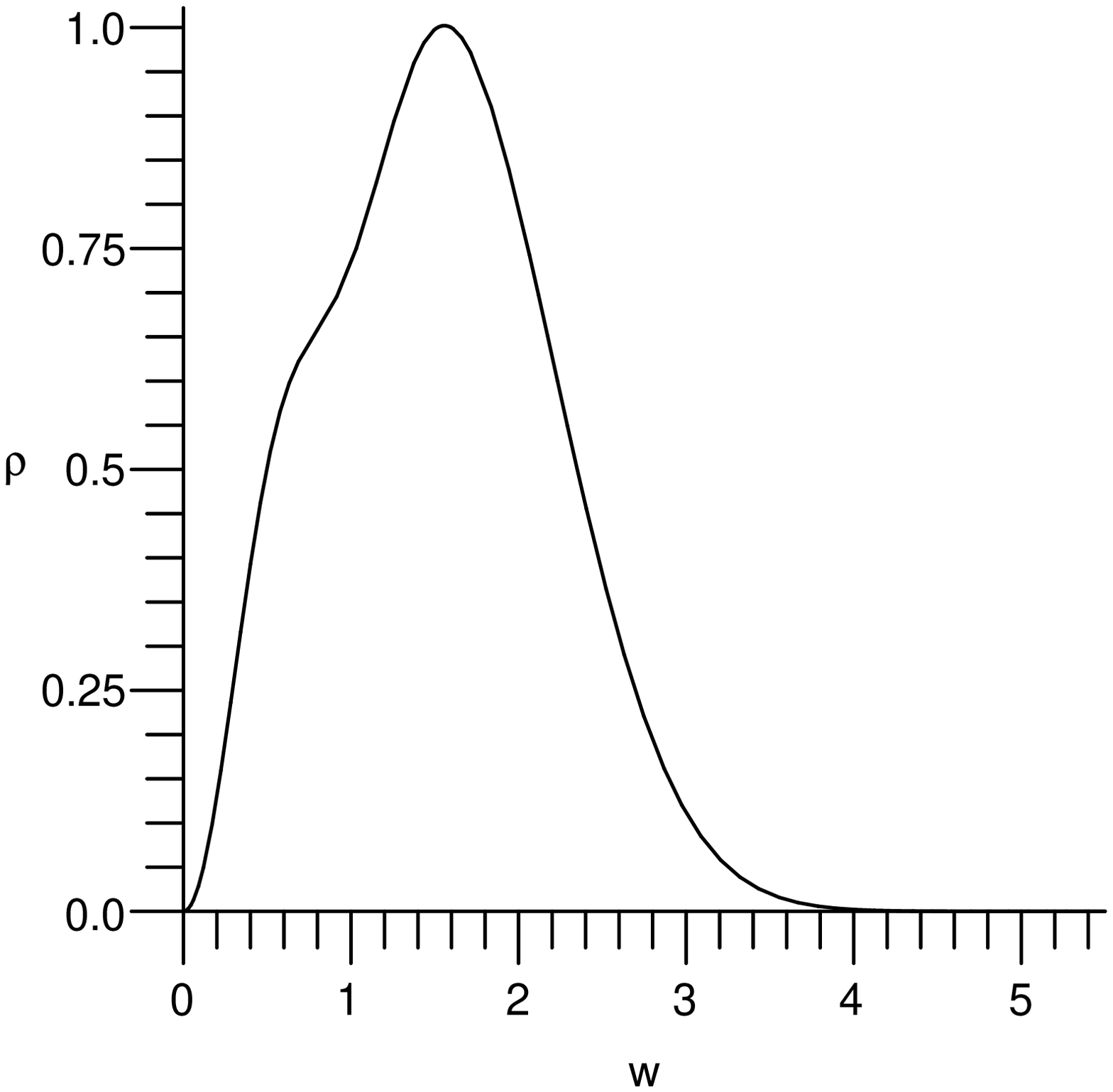}
\caption{
The probability density for the wave packet given in Eq. (\ref{wpsi}) for $T = 1.04$.}
\end{figure}
\vskip 10 pt

\begin{figure}
\includegraphics[width=5in]{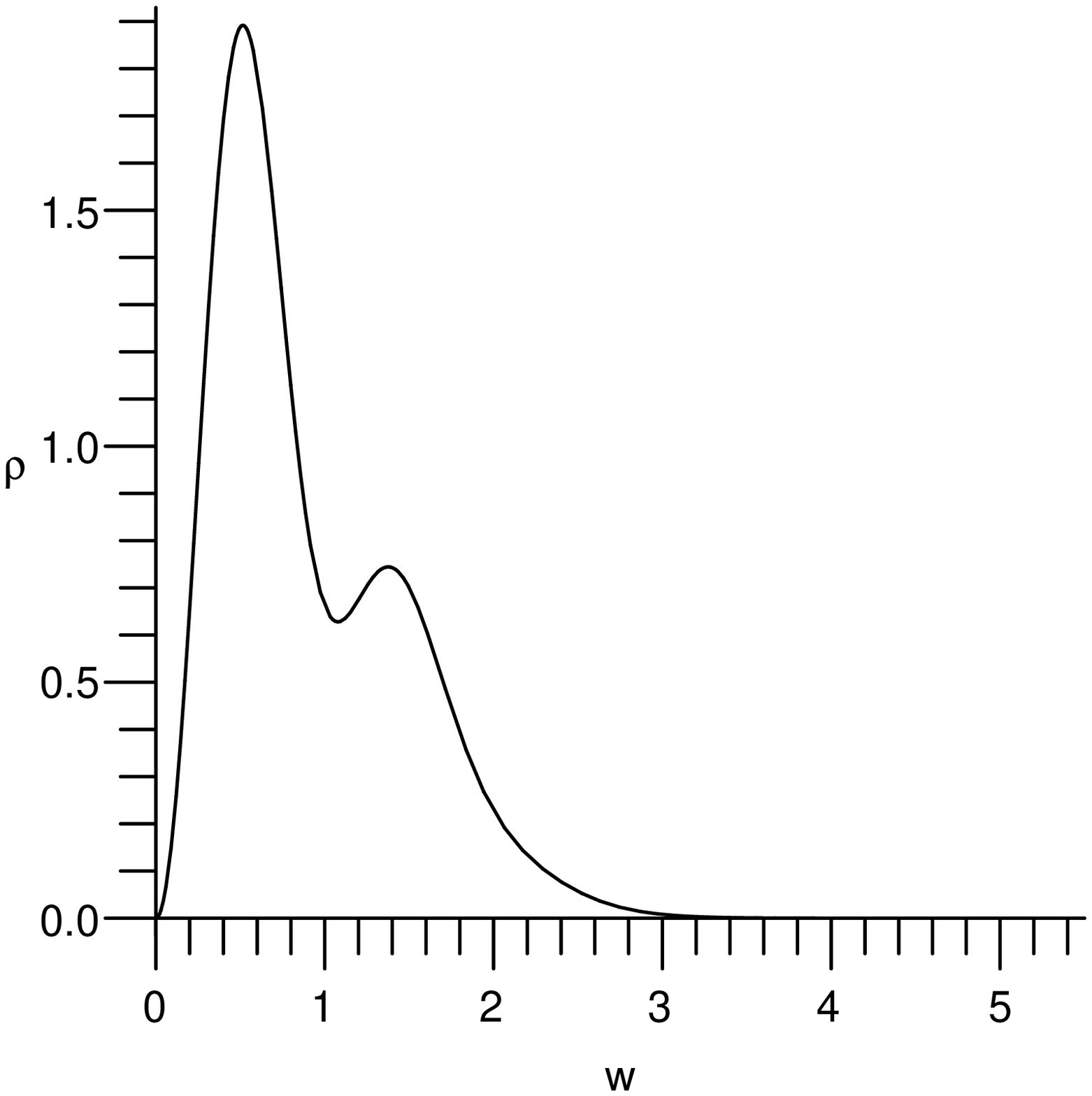}
\caption{
The probability density for the wave packet given in Eq. (\ref{wpsi}) for $T = 1.3$.}
\end{figure}
\vskip 10 pt

\begin{figure}
\includegraphics[width=5in]{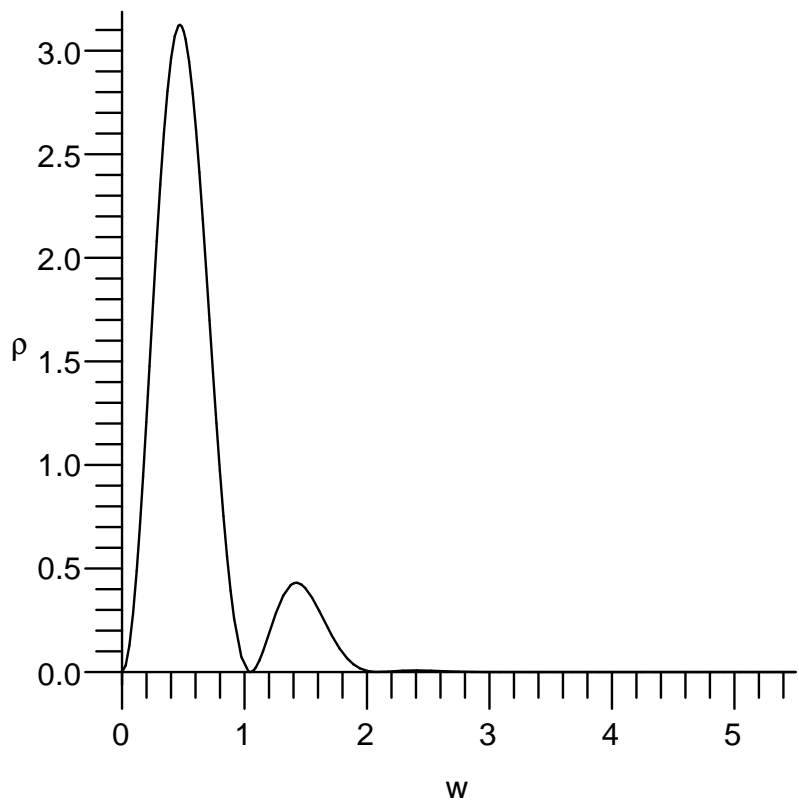}
\caption{
The probability density for the wave packet given in Eq. (\ref{wpsi}) for $T = 1.57$, $\approx \pi/2$.}
\end{figure}
\vskip 10 pt

\begin{figure}
\includegraphics[width=5in]{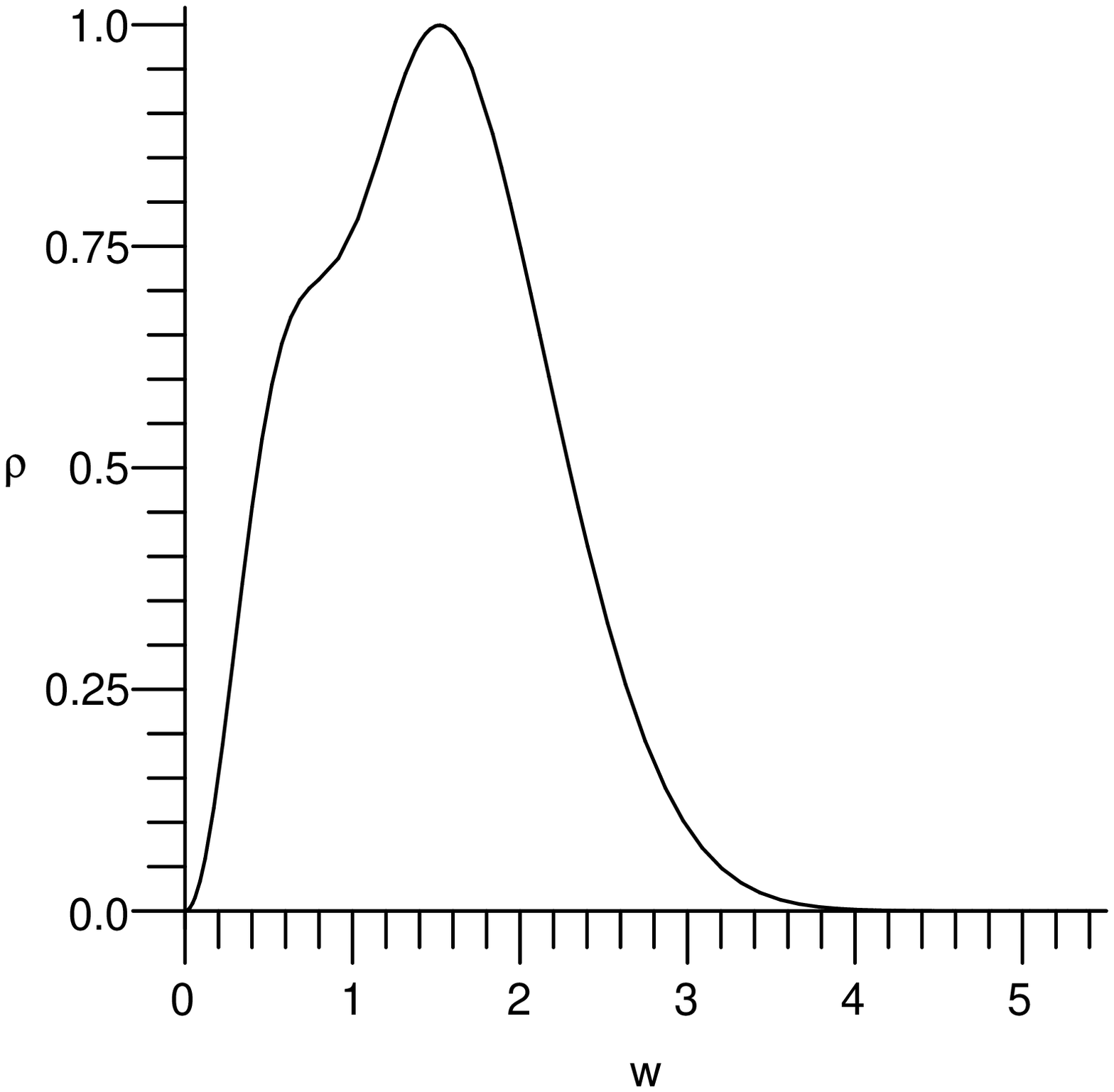}
\caption{
The probability density for the wave packet given in Eq. (\ref{wpsi}) for $T = 2.08$.}
\end{figure}
\vskip 10 pt

\begin{figure}
\includegraphics[width=5in]{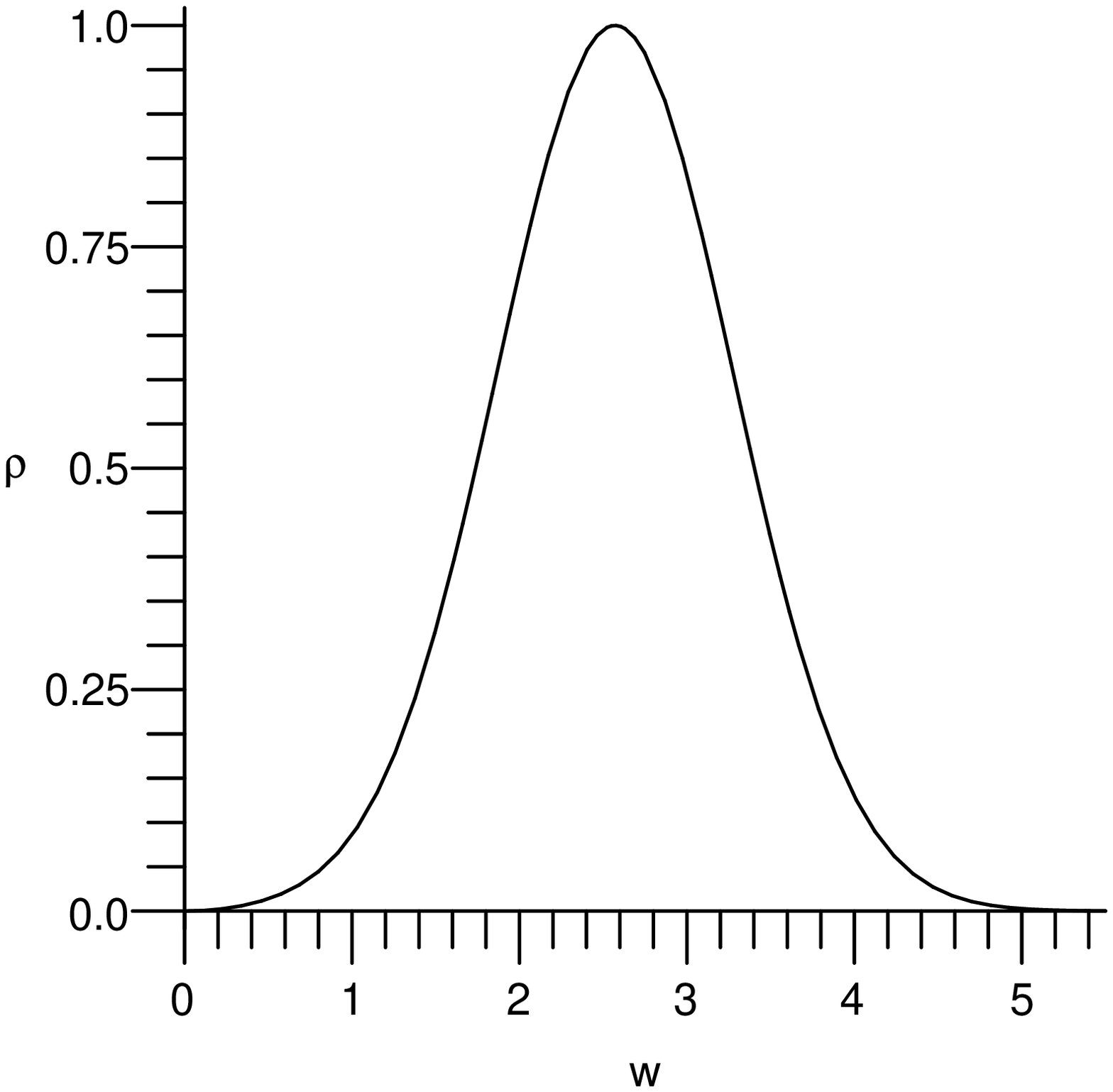}
\caption{
The probability density for the wave packet given in Eq. (\ref{wpsi}) for $T = 2.6$.}
\end{figure}
\vskip 10 pt

\begin{figure}
\includegraphics[width=5in]{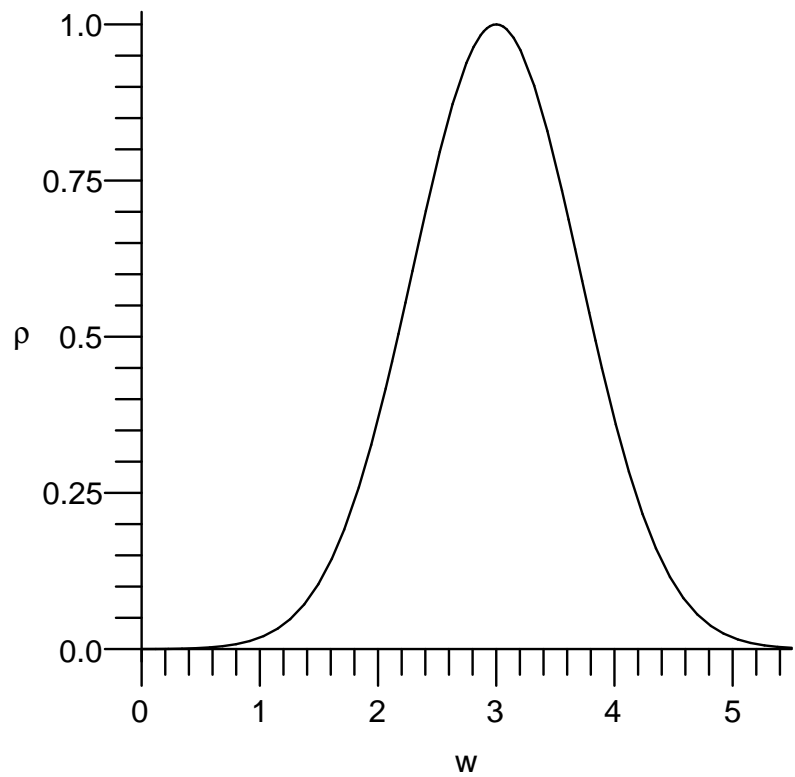}
\caption{
The probability density for the wave packet given in Eq. (\ref{wpsi}) for $T = 3.14$, $\approx \pi$.}
\end{figure}
\end{document}